\documentclass[aps, prl, twocolumn,preprintnumbers, nofootinbib, showpacs, superscriptaddress,letterpaper]{revtex4}
\usepackage{amsmath}
\usepackage{amsfonts}
\usepackage{amssymb,multirow,wrapfig}
\usepackage{color}
\usepackage{graphicx}  
\usepackage{cancel}
\usepackage{adjustbox}
\usepackage{array}
\usepackage{booktabs}
\usepackage{multirow}
\usepackage{lineno}
\usepackage{dsfont}
\usepackage{xspace}
\usepackage{float}

\newcommand{\bra}[1]{\ensuremath{\left< #1 \right|}}
\newcommand{\ket}[1]{\ensuremath{\left| #1 \right>}}

\newcommand{\ketbra}[2]{\ensuremath{\left| #1\left>  \right<#2 \right|}}
\newcommand{\avg}[1]{\ensuremath{\left< #1 \right>}}

\newcommand{\be}[0]{\begin{equation}}
\newcommand{\ee}[0]{\end{equation}}
\newcommand{\bea}[0]{\begin{eqnarray}}
\newcommand{\eea}[0]{\end{eqnarray}}

\newcommand{\expval}[1]{\left\langle #1 \right\rangle}
\newcommand{\sig}[3]{\sigma_{\text{#1}}^{\text{#2},\text{#3}}}
\newcommand{\melvin}{{\small M}{\scriptsize ELVIN}\xspace}

\begin{document}
\title{Experimental GHZ Entanglement beyond Qubits}

\author{Manuel Erhard}
\email{manuel.erhard@univie.ac.at}
\affiliation{Institute for Quantum Optics and Quantum Information (IQOQI), Austrian Academy of Sciences, Boltzmanngasse 3, A-1090 Vienna, Austria}
\affiliation{Faculty of Physics, University of Vienna, Boltzmanngasse 5, 1090 Vienna, Austria}

\author{Mehul Malik}
\affiliation{Institute for Quantum Optics and Quantum Information (IQOQI), Austrian Academy of Sciences, Boltzmanngasse 3, A-1090 Vienna, Austria}
\affiliation{Faculty of Physics, University of Vienna, Boltzmanngasse 5, 1090 Vienna, Austria}

\author{Mario Krenn}
\affiliation{Institute for Quantum Optics and Quantum Information (IQOQI), Austrian Academy of Sciences, Boltzmanngasse 3, A-1090 Vienna, Austria}
\affiliation{Faculty of Physics, University of Vienna, Boltzmanngasse 5, 1090 Vienna, Austria}

\author{Anton Zeilinger}
\email{anton.zeilinger@univie.ac.at}
\affiliation{Institute for Quantum Optics and Quantum Information (IQOQI), Austrian Academy of Sciences, Boltzmanngasse 3, A-1090 Vienna, Austria}
\affiliation{Faculty of Physics, University of Vienna, Boltzmanngasse 5, 1090 Vienna, Austria}

\date{\today}

\begin{abstract} 
The Greenberger-Horne-Zeilinger (GHZ) argument provides an all-or-nothing contradiction between quantum mechanics and local-realistic theories. In its original formulation, GHZ investigated three and four particles entangled in two dimensions only. Very recently, higher dimensional contradictions especially in three dimensions and three particles have been discovered but it has remained unclear how to produce such states. In this article we experimentally show how to generate a three-dimensional GHZ state from two-photon orbital-angular-momentum entanglement. The first suggestion for a setup which generates three-dimensional GHZ entanglement from these entangled pairs came from using the computer algorithm \melvin. The procedure employs novel concepts significantly beyond the qubit case. Our experiment opens up the possibility of a truly high-dimensional test of the GHZ-contradiction which, interestingly, employs non-Hermitian operators.
\end{abstract}

\maketitle
\section{Introduction}
Developed in the early 20th century, quantum mechanics forms the basis of our modern understanding of microscopic physics. Many technological developments such as the laser and semi-conductors are based on the principles of quantum mechanics. In addition to its wide usage in different fields of science and technology nowadays, quantum theory has spurred profound questions about the nature of reality itself. The concept of reality refers to the fact that physically measurable quantities of objects possess their properties independent of and prior to a measurement. The principle of locality states that in order to ensure a causal structure of the universe, physical influences must propagate at a finite speed (the speed of light). The question whether quantum mechanics describes a universe that is local and realistic was first raised in 1935 by Einstein, Podolsky and Rosen, who considered a pair of non-separable or entangled quantum particles~\cite{einstein1935can}. Almost 30 years later, John Bell elevated this meta-physical discussion to an experimentally testable theorem~\cite{bell1964einstein} that has found ever-increasing experimental support in recent years~\cite{hensen2015loophole,giustina2015significant,shalm2015strong}.

In close analogy to Bell's theorem, Greenberger, Horne, and Zeilinger (GHZ)~\cite{greenberger1989going,greenberger1990bell} found that the entanglement of at least three particles results in a stronger violation of local-realism. In contrast to Bell's theorem that tests whether statistical predictions of quantum mechanics are in conflict with local-realism, the GHZ argument uses definite (non-statistical) predictions to show a contradiction between quantum theory and local-realistic theories. While the original GHZ argument for two-dimensional entangled states was formulated more than two decades ago~\cite{greenberger1989going,mermin1990extreme}, its generalization to states of arbitrary dimensions was found only recently~\cite{ryu2013greenberger,PhysRevA.89.024103,lawrence2014rotational,tang2017multisetting}. Likewise, experimental studies of GHZ entanglement have thus far only considered quantum states in two dimensions, i.e. qubits. GHZ states have been realized in a diverse range of physical systems, including 14 ions~\cite{monz201114}, up to 10 photons~\cite{bouwmeester1999observation,jian2000experimental,PhysRevLett.117.210502} and 10 super-conducting qubits~\cite{kelly2015state,song201710}. In all of these systems, a general recipe exists for increasing the number of entangled particles. 

However, there is no such recipe for extending the dimensionality of each individual entangled particle beyond qubits. A possible theoretical recipe would be to use a generalization of the CNOT gate to higher dimensions~\cite{wilmott2011swapping}. However, the question of how to experimentally realize such a gate remains an open one. To find a direct realization, we tried utilizing the symmetries of the state and analogies to the qubit experiments. Unfortunately, all of these approaches did not prove to be fruitful. Using an alternative approach, we developed a computational algorithm called \melvin to search for possible experimental realizations~\cite{krenn2016automated}. \melvin has no intuition or prefixed idea of how such an experiment could look. Therefore, this computer algorithm was able to find surprisingly resource-efficient but for us humans counter-intuitive experimental realizations of such complex quantum states.

In this article we show the experimental creation of the first three-particle Greenberger-Horne-Zeilinger state in three dimensions. The experimental scheme presented here is one of several new experiments proposed and inspired by our computer algorithm \melvin~\cite{krenn2016automated,schlederer2016cyclic,babazadeh2017high,PhysRevLett.118.080401}. Furthermore, we use an entanglement witness to show that our generated state is indeed genuinely three-particle and three-dimensionally entangled. Additionally, we propose an experimental scheme for a multi-setting, high-dimensional multi-partite violation of local-realism using our entangled state.
\begin{figure*}[ht]
	\centering
	\includegraphics[width=0.8\textwidth]{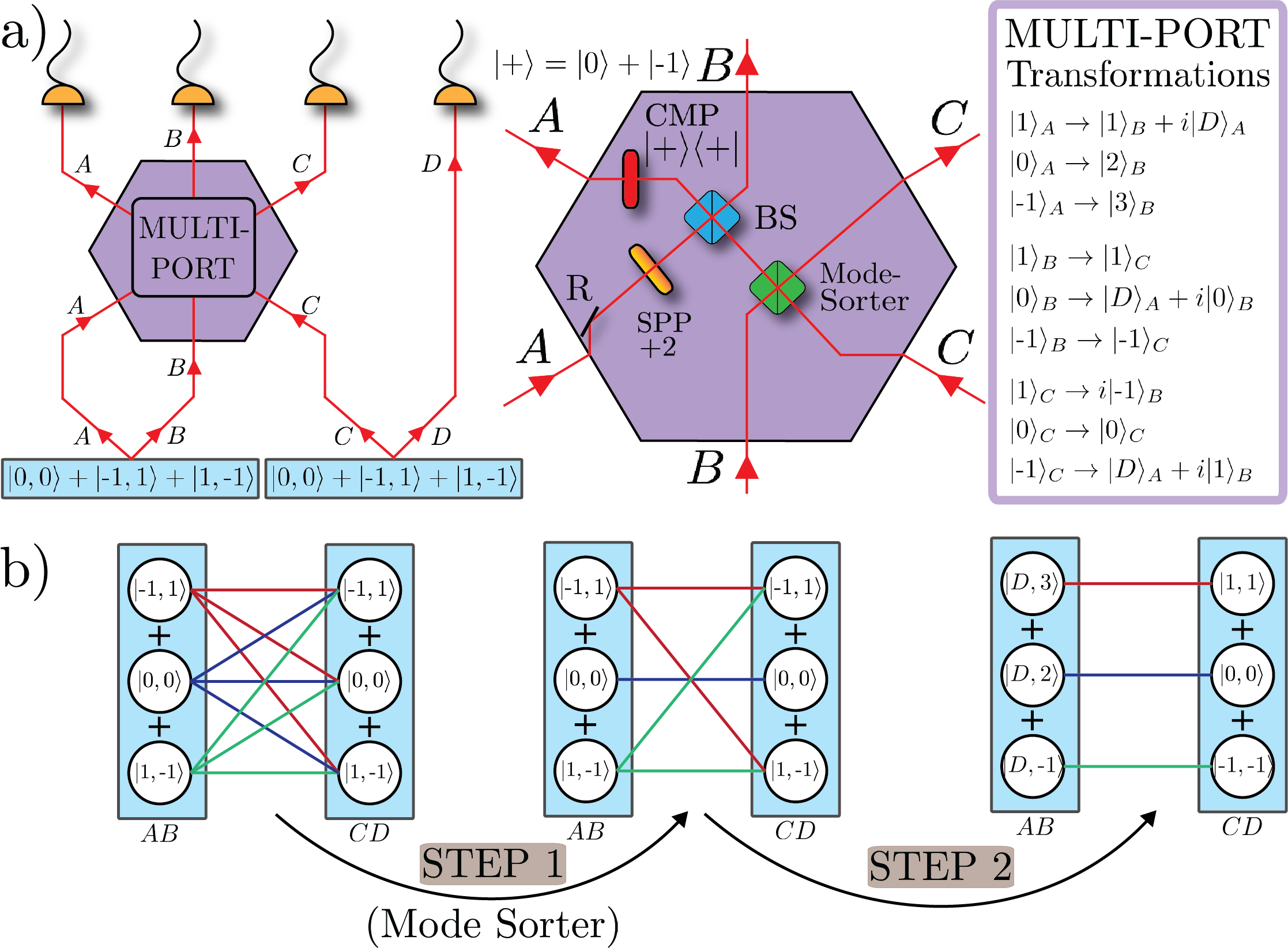}
	\caption{\textbf{Schematic of the experiment.} \textbf{a)}Two non-linear crystals (NLC) each produce a three-dimensionally entangled pair of photons. A high-dimensional multi-port (MP) transforms each photon according to the transformations displayed above. The experimental implementation of the MP is shown in the inset. An orbital-angular-momentum mode (OAM) sorter (depicted in green) sorts incoming photons according to their OAM value (even/odd). A reflection (R) in combination with a spiral-phase-plate (SPP) in path A changes the OAM value from $\ket{\ell}\rightarrow\ket{\text{-}\ell+2}$. A beam-splitter (BS) coherently combines paths A and B. Finally, a coherent-mode-projection (CMP) projects photons in path A onto $\ket{+}=\ket{0}+\ket{\text{-1}}$.\\
	\textbf{b)} \textbf{Physical principle behind the creation of a three-dimensional GHZ state.} Each NLC coherently emits a pair of high-dimensionally entangled photons. The overall four-photon probability amplitudes ($3\times 3=9$) are represented by the red, green and blue lines. In step 1, the OAM mode-sorter inserted in path B and C prevents a four-fold detection event between even and odd terms emitted by different crystals. In step 2, the multi-port eliminates the remaining two cross-connections between the two crystals. Thus only three four-photon probability amplitudes are left. All photons exiting path A of the multi-port are in the $\ket{+}$ state. Thus the final three-dimensional GHZ state is created in the three paths $B,C~\text{and}~D$ and reads $\ket{3,1,1}+\ket{2,0,0}+\ket{\text{-}1,\text{-}1,\text{-}1}$.}
	\label{fig:exp-scheme}
\end{figure*}
\section{Experiment}

We choose the orbital-angular-momentum (OAM) of photons~\cite{allen1992orbital,yao2011orbital,krenn2017orbital} as a physical carrier of information in our experiment. The OAM of photons spans an in-principle infinite-dimensional, discrete state space and can thus easily encode three different quantum levels. A three dimensional, three-particle GHZ state is written as:
\begin{equation*}
	\ket{\psi}=(\ket{0}_a\ket{0}_b\ket{0}_c+\ket{1}_a\ket{1}_b\ket{1}_c+\ket{2}_a\ket{2}_b\ket{2}_c)/\sqrt{3},
\end{equation*} where the numbers refer to the three different states and the letters to different particles. In this representation, the perfect correlations among the GHZ state are clearly visible. If one of the particles, say $a$ is measured to be in state $\ket{2}$, then particles $b,c$ are also found in state $\ket{2}$. Despite these simple correlations, it is a challenging task to create such a state experimentally. Below, we describe the at-first-sight counterintuitive approach found by our computer algorithm \melvin to create such a state.

\begin{figure}[htbp]
	\includegraphics[width=0.45\textwidth]{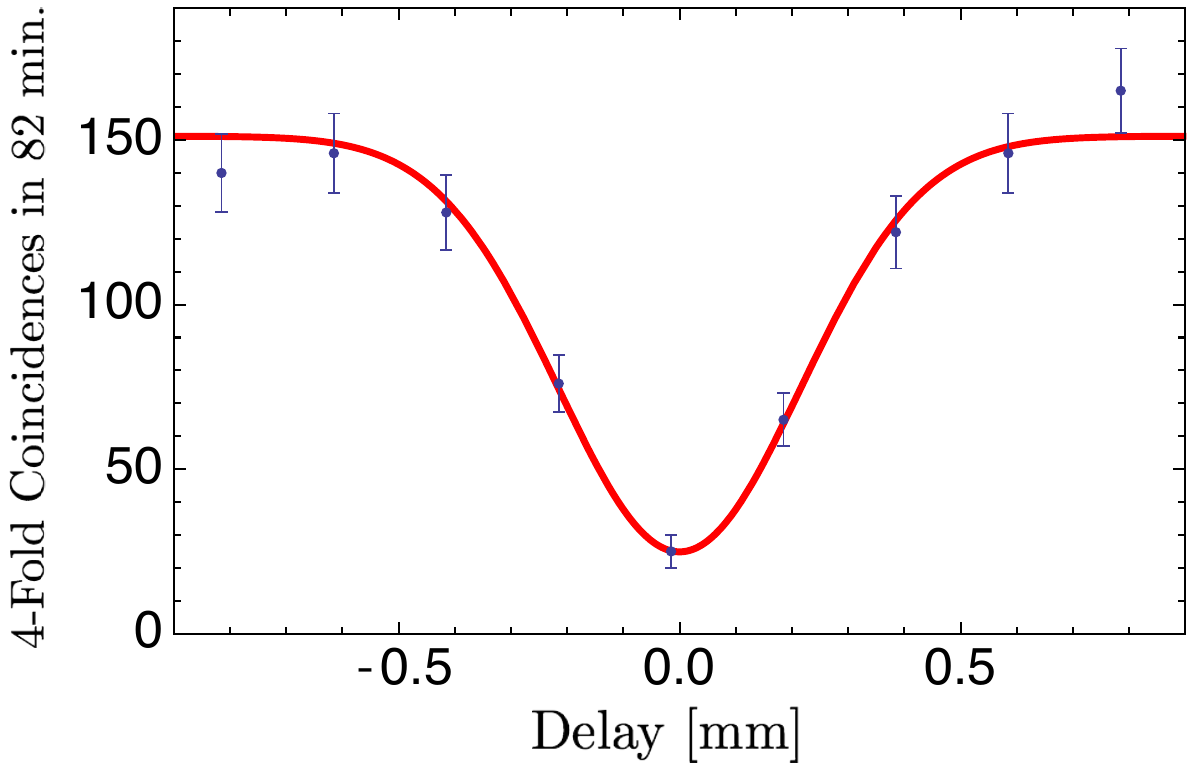}
	\caption{\textbf{Four-Photon Hong-Ou-Mandel Interference.} Experimental data showing the four-photon HOM effect by measuring $\ketbra{-1}{-1}_A\otimes\ketbra{1}{1}_B\otimes\ketbra{-1}{-1}_C\otimes\ketbra{1}{1}_D$. The drop in four-fold ``click'' events is due to the impossibility that photon $A$ and $B$ exit the beam-splitter in different paths if they are indistinguishable.  Fitting the experimental data points with an assumed Gaussian spectrum yields a visibility of $83.4\%$ and a width of $800\mu \text{m}$. Error bars indicate Poissonian noise in the photon-count rate.}  
	\label{fig:hom-1}
\end{figure}
 
We start with two non-linear crystals that each produce a three-dimensionally entangled state of two photons. As shown in Fig.~\ref{fig:exp-scheme}, each non-linear crystal creates two probability amplitudes containing odd OAM units ($\ket{\text{-}1,1}+\ket{1,\text{-}1}$) and one probability amplitude with even units of OAM ($\ket{0,0}$). If both crystals simultaneously and coherently emit a photon pair, then $3\times 3=9$ different four-photon probability amplitudes occur. All combinations are represented by red, green and blue lines connecting the different probability amplitudes emitted by the two crystals in Fig.~\ref{fig:exp-scheme}b). The goal is that only three of these probability amplitudes remain in the final three-dimensional GHZ state. Thus, we need a clever way to prevent the other six probability amplitudes from occurring in our experiment. First, we only focus on cases in which a photon is simultaneously detected in each detector A, B, C and D, comprising a four-fold detection event. In addition, we use a new type of multi-port that operates on a high-dimensional state-space with three input and three output ports.

In comparison with a polarizing beam splitter that is routinely used to create two-dimensional GHZ states in polarization, the high-dimensional multi-port (MP) has several special features. One of these features is that for every input port $(A,B,C)$, a different state from the basis set $\{\ket{-1},\ket{0},\ket{1}\}$ is transformed into a coherent superposition of two output ports, as shown in Fig.~\ref{fig:exp-scheme}b). This will turn out to be a key feature of the multi-port. From the transformation table for the multi-port shown in Fig.~\ref{fig:exp-scheme}, we can now understand how the three-dimensional GHZ state is generated in our experimental setup.

The transformation rules of inputs B and C show that a four-fold event can only occur if both crystals emit either even or odd-valued two-photon OAM probability amplitudes. All mixtures between even and odd states are thus prevented by the OAM mode-sorter, comprising the first step in the multi-port. The working principle of the mode-sorter itself is based on interferometry~\cite{leach2002measuring} and it coherently sorts OAM values according to their parity and thus also conserves the entanglement between the respective modes~\cite{erhard2017quantum}. Therefore, only five possibilities remain which all have the same parity, see Fig.~\ref{fig:exp-scheme} b). For the generation of a two-dimensional GHZ state, such a distinction is already sufficient to create a GHZ-like entangled state. In the three-dimensional case, however, there is still the cross-correlation between the odd OAM probability amplitudes, as shown in Fig. 1b.

The possibility $\ket{-1,1}_{AB}\otimes\ket{1,-1}_{CD}$ is prevented by the multi-port because it does not result in a four-photon detection event after the transformations shown in Fig. 1a. Interestingly, the other cross-connection between $\ket{1,-1}_{AB}\otimes\ket{-1,1}_{CD}$ is prevented by two-photon interference at the beam splitter (BS), the so-called Hong-Ou-Mandel (HOM) effect~\cite{hong1987measurement}. This effect can be directly measured and the experimental result is shown in Fig.~\ref{fig:hom-1}. Ideally, the probability of simultaneously measuring a photon in each detector is zero. However, this probability depends on the distinguishability of the photons involved (see Supplementary for details). Therefore, this measurement also serves to estimate how well the multi-port and the two entanglement sources work. In our case, the probability amplitude $\ket{1,-1}_{AB}\otimes\ket{-1,1}_{CD}$ is suppressed by 83.4\%.

After the two steps through the multi-port, we are left with three remaining links. These connections represent the generated three-dimensional GHZ state. The photon in path A is always in state $\ket{+}=\ket{0}+\ket{\text{-}1}$ state and can therefore be factorized from the other three photons B,C and D. This means that photon A is no longer entangled with the other three photons. The remaining probability amplitudes undergo a transformation according to the transformation rules imposed by the multi-port. Thus the final state created in paths $B,C,D$ reads:
\begin{equation*}
	\ket{\psi}=\frac{1}{\sqrt{3}}(\ket{2,0,0}+\ket{3,1,1}+\ket{-1,-1,-1}).
\end{equation*}
This state is exactly the desired three-dimensional three-particle GHZ state, which can be seen by locally changing the modes $2\rightarrow 0$ and $3\rightarrow 1$.

\begin{figure}[htbp]
	
	\includegraphics[width=0.45\textwidth]{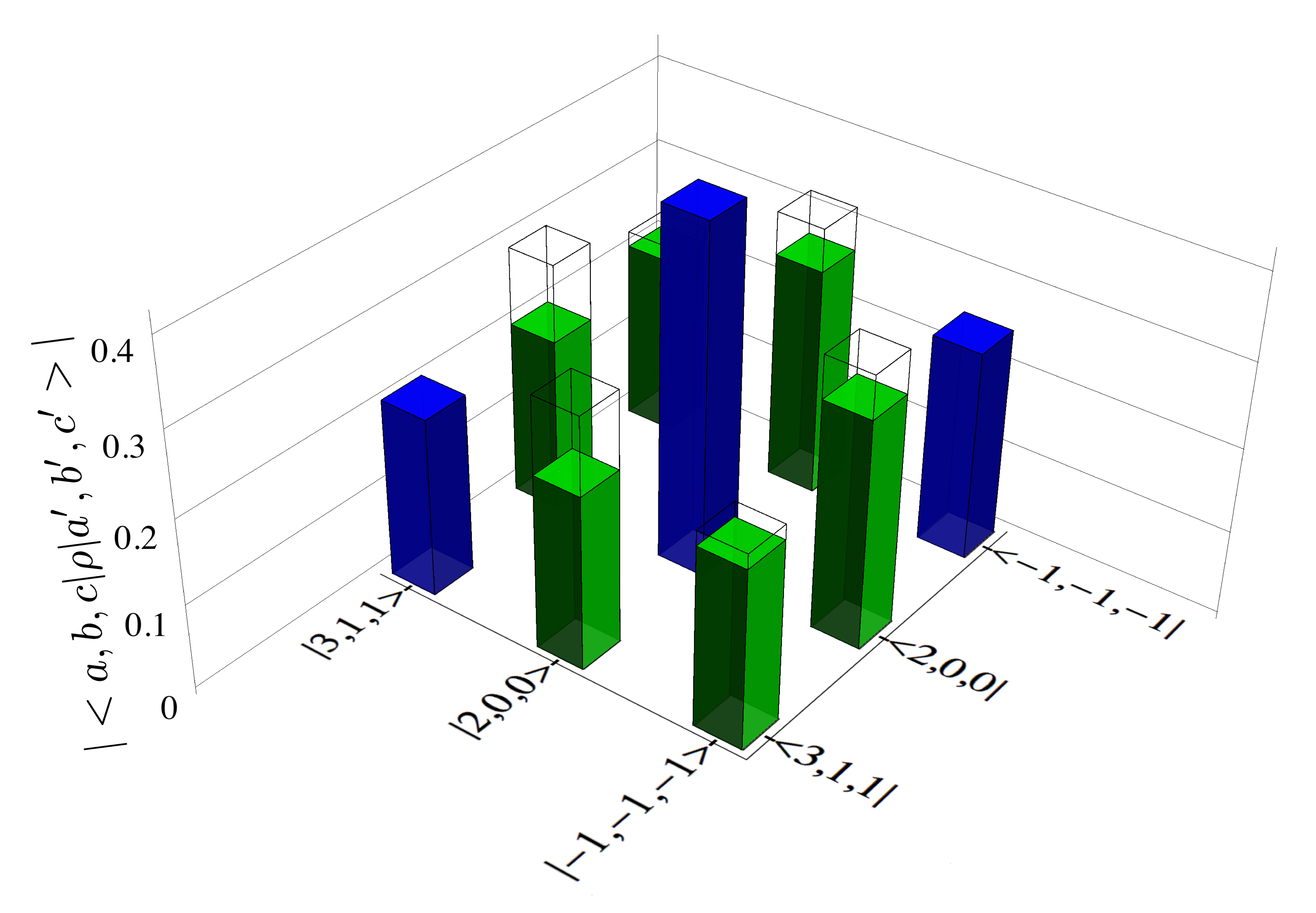}
	\caption{\textbf{Experimentally measured density matrix elements.} Here the measured density matrix elements for calculating the fidelity $\text{F}_\text{exp}=75.2\%$ are depicted (blue/green elements represent diagonal/off-diagonal entries). This verifies genuine multi-partite entanglement in (3,3,3) dimensions with 3 standard deviations. The unbalancing of the diagonal elements results from the three-dimensional two-photon sources and is expected. Furthermore, $87.8\%$ of the detected counts of the diagonal elements are in the expected elements. The average coherence of the measured state is approximately $81.7\%$. Perfect coherence is indicated by empty bars.}
	\label{fig:exp-density-matrix}
\end{figure}

We use an entanglement-dimension witness~\cite{malik2016multi} to verify that our three-photon state is indeed genuinely multi-partite entangled in three dimensions. This approach is based on the idea that the overlap of an ideal three-dimensional GHZ state with any state from a lower dimensional entanglement structure cannot exceed a certain maximum value. If our measured state exceeds this maximum fidelity, it is genuinely multi-partite entangled in dimension three. The entanglement structure is defined according to the Schmidt-Rank-Vector (SRV) formalism~\cite{huber2013structure}. Each number in the SRV corresponds to the entanglement dimensionality of one party with respect to the remaining two parties. Thus for the GHZ state, all three bi-partitions $\{A|BC,~B|AC,~C|AB\}$ are three-dimensionally entangled, giving an $\text{SRV}=(3,3,3)$. The maximum possible fidelity between a $(3,3,3)$ state $\ket{\psi}$ and any quantum state $\chi$ with a smaller dimensionality structure, e.g. $\chi\in(3,3,2)$ is $\text{F}_\text{max}=\max\limits_{\chi\in(i,j,k)} \text{Tr}(\chi \ket{\psi}\bra{\psi})\leq2/3$, for all $\text{permutations of}~(i,j,k)~\text{with}~i,j\leq 3~ \text{and}~ k\leq 2$. Thus if the fidelity of our experimentally created state $\rho$, $\text{F}_\text{exp}=\text{Tr}(\rho\ket{\psi}\bra{\psi})$ exceeds this bound $\text{F}_\text{max}$, we have shown that we have indeed created a genuinely $(3,3,3)$-dimensionally entangled state. This witness has the remarkable advantage that only the expected non-zero density matrix elements need to be measured, which significantly reduces the measurement time. 

The absolute values of the measured density matrix elements are depicted in Fig.~\ref{fig:exp-density-matrix}. The diagonal elements are simple projection measurements in the computational basis. However, each off-diagonal element is reconstructed from 64 consecutive two-dimensional subspace measurements. Hence, a total of 219 measurements are performed in order to reconstruct the necessary density matrix elements. In total, we observed 1652 simultaneous four-photon ``click'' events in 378 hours. Due to the long measurement time, we also applied accidental-count subtraction (see Supplementary for details). From this data, we calculate the experimental fidelity to be $\text{F}_\text{exp}=75.2\%\pm2.88\%$ which certifies with 3 standard deviations that the observed state is indeed genuinely three-dimensional and three-photon entangled. The error was calculated using Monte-Carlo simulations with Poissonian counting statistics.

\section{Multi-Setting Three-Particle GHZ Experiment in Three-Dimensions}
\begin{figure*}[htbp]
	\centering
	\includegraphics[width=0.95\textwidth]{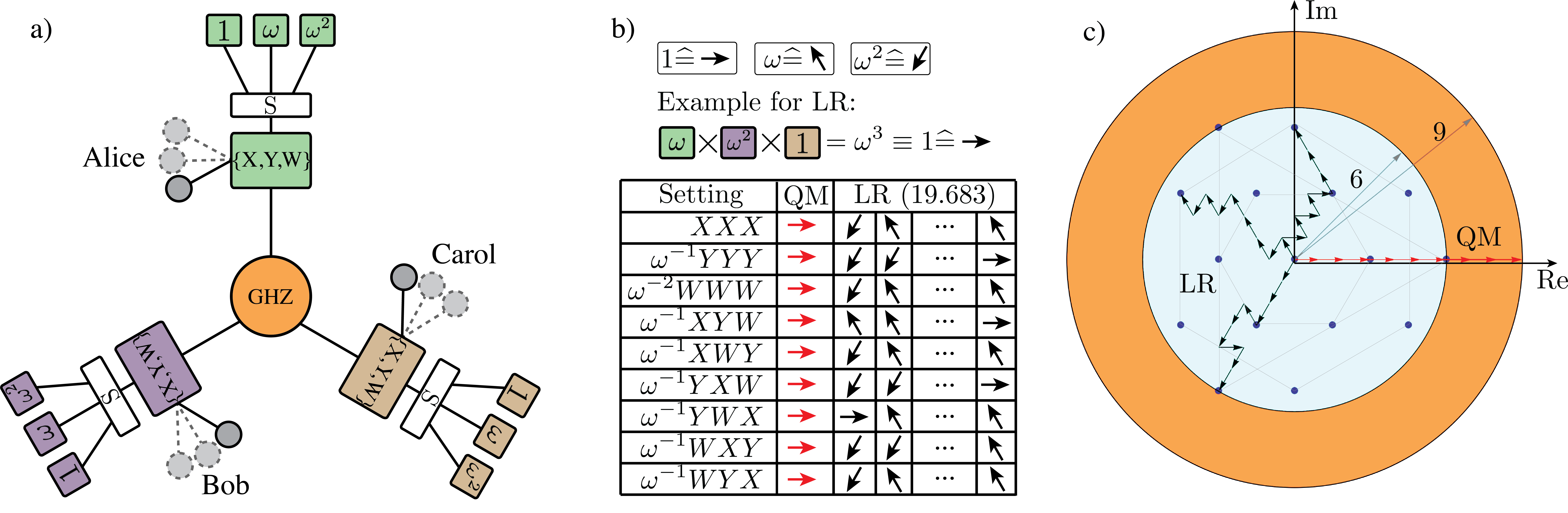}
	\caption{\textbf{a)~Schematic of the GHZ contradiction.} The three-particle and three-dimensionally entangled GHZ state is produced in the center and then send to Alice, Bob and Carol. They are spatially separated and can independently choose one of the three transformations $\{X,Y,W\}$. An orbital angular momentum sorter (S) then routes the different OAM values $\ell\in\{0,1,2\}$ into different paths. Each path is now assigned with a number corresponding to the OAM value, such that $v(\ell)=\omega^\ell$. \textbf{b)} Each value $v(\ell)$ can now be identified with an arrow of unit length and a corresponding direction. Since these values form a group under multiplication, the multiplication results again in one of these values. Quantum mechanically, all different settings result in precisely the same outcome, namely 1. This shows the perfect correlations present in quantum mechanics. Local-realistically there exist 19.683 different possibilities, but non of them is perfectly correlated. \textbf{c)} If we sum up the results from the nine settings, we see quantum mechanically we can reach a value of 9. In the local-realistic picture, there exist sixteen possible values (shown as blue dots), nicely arranged on regular convex polygons. But each of them is contained within a circle of radius 6. This clearly shows the contradiction between quantum mechanics and local-realism.}
	\label{fig:proposed-ghz-experiment}
\end{figure*}

Although the extension from two to three-dimensional GHZ states might seem to be only a modest step forward, the implications are rather profound. Our results confirm that an experiment is possible where the GHZ contradiction is realized in higher-dimensions. Such a contradiction has novel interesting features beyond the qubit case. Here we follow the theoretical construction given by Lawrence~\cite{lawrence2014rotational,lawrence2017mermin} and explicitly point to the conceptual differences as compared to the qubit case. Most interestingly to use hermitian local observables does not lead to a GHZ contradiction. Likewise the three-particle observables do not commute as they do in the qubit case. It turns out that the crucial feature which is retained in the three-dimensional case is the unitarity of all observables.

The essential ingredients for an all-versus-nothing violation of local-realistic theories are that we use single-particle observables and that the observables have definite outcomes. This means that in the quantum mechanical description, the three-body observables are constructed from the tensor product of the local observables and that all these observables have the GHZ state as a common eigenstate. A set of such observables is called a concurrent set~\cite{PhysRevA.73.032316}. These three-body observables don't necessarily need to commute, as they do in the two-dimensional case. Also in contrast to the original two-dimensional GHZ proposal, we use three local observables called $X, Y, \text{and}~ W$. Since $X,Y,~\text{and}~W$ are unitary operators, their eigenvalues are complex numbers. This is in stark contrast to the 2-dimensional case where hermitian observables with real eigenvalues are used. In analogy to the qubit case we define a Mermin operator $\mathcal{O}$ (see Supplementary for details) which is maximized by the GHZ state and yields a value of 9. Under local-realistic assumptions, the maximum value for the Mermin operator is 6, as shown in Fig.~\ref{fig:proposed-ghz-experiment}c). Thus in a real experiment, one can only observe a value greater than 6 by violating either the \textit{locality} or the \textit{element of reality} assumption. 

In our proposed experiment, the photons are sent to three spatially separated observers, see Fig.~\ref{fig:proposed-ghz-experiment}~a). Alice, Bob and Carol then have the possibility to set one of the three measurement settings X, Y and W at their local measurement device. This measurement apparatus consists of a unitary transformation, which transforms the photon locally into the eigenbasis of the measurement setting. Such transformations have recently been found by our computer algorithm \melvin and also experimentally demonstrated~\cite{babazadeh2017high,wang2017generation}. Subsequently, a Stern-Gerlach type apparatus sorts the photons according to their OAM value $\ket{0},\ket{1}$ or $\ket{2}$ \cite{leach2002measuring,berkhout2010efficient,mirhosseini2013efficient}. In each of these three paths, a single-photon detector is assigned to a complex number, namely $\{1,\omega,\omega^2\}$. Decisive is now the conditional probability to consider when the three observers at the same time receive a certain measurement result. A brief example to illustrate: Alice, Bob and Carol all select the setting X at their local measuring devices. Then all of them get a measurement result which gives one of the three possible values ​​$\{1,\omega,\omega^2\}$ with equal probability. If, however, the three local measurement results of Alice, Bob and Carol are multiplied, the result is always one, as mentioned above. This reflects the perfect correlations which are present in the GHZ state.

Such perfect correlations are only possible with an ideal state. This is difficult to achieve in a real experiment. But perfect correlations are not necessary for a statistically significant violation of local-realistic theories. The decisive criteria for the quality of the state produced and whether a violation is possible in principle is discussed here. The three criteria which determine the quality of the generated state are white-noise, average coherence and weighting of the individual terms in the GHZ state. From our experimental data we see that on average, the expected versus the observed magnitude of the off-diagonal elements of our state is $81.7\%$, which therefore measures the average coherence. Additionally, $87.8\%$ of the detected counts in the diagonal elements are in the expected elements, thus $12.2\%$ is the amount of white-noise present in our state. 
We then theoretically construct a density matrix $\rho_{\text{p}}$ which contains these three parameters and calculate from this the expectation value for the generalized Mermin operator $\mathcal{O}$, which yields with our parameters a result of $\avg{\mathcal{O}}_{\rho_{\text{p}}}=6.26\pm0.25$ (details in the Supplementary). The limit for local-realistic theories is 6. It is therefore realistic that with our experimentally generated state such an experiment can actually be carried out. Realistic improvements could include novel high-dimensional two-photon sources~\cite{PhysRevLett.118.080401} to achieve an equal weighting of the individual terms in the GHZ state. In addition, it is possible to significantly increase the photon counting rate with this method. This makes it possible to use narrower spectral filters, which in turn improves the average coherence.

\section{Conclusion}
In conclusion, we have shown the first experimental realization of a Greenberger-Horne-Zeilinger state entangled in three dimensions for each of the three photons. Remarkably, our experimental method for generating this state was found through the use of a computer algorithm called \melvin. The unconventional approach proposed by \melvin not only made the actual generation of this state possible, but also shows that such algorithms are a key resource for finding ways to create increasingly complex quantum states. In addition, we have elaborated a three-dimensional multi-setting ``all-or-nothing" test of local-realism that can be implemented with our entangled state. For such experiments, an open theoretical question remains as to whether one can find stronger and more robust violations of local-realistic theories for multi-partite and high-dimensional entanglement than the one presented here. On the applications front, this work opens new opportunities to experimentally investigate quantum-secret-key sharing in higher dimensions. Such protocols based on higher-dimensional GHZ states offer an unprecedented level of security against several different hacking attacks~\cite{bai2017quantum}. Another interesting application is to utilize high-dimensional encoding to reduce the complexity of quantum computing algorithms~\cite{bocharov2016factoring}.

\section{Acknowledgments}
We thank Jay Lawrence, Marcus Huber, Caslav Brukner, Armin Hochrainer, Robert Fickler, Thomas Scheidl and Fabian Steinlechner for fruitful discussions. This work was supported by the Austrian Academy of Sciences (\"OAW), by the European Research Council (SIQS Grant No. 600645 EU-FP7-ICT) and the Austrian Science Fund (FWF) with SFB F40 (FOQUS) and FWF project CoQuS No. W1210-N16. M.M. would like to acknowledge support from the European Commission through a Marie Curie fellowship (OAMGHZ), the Austrian Science Fund (FWF) through the START project Y879-N27, and the joint FWF-GACR project MULTIquest.

\bibliography{3d-ghz.bib}
\newpage
\newpage
\section{Supplementary}
\subsection{Three-Dimensionally Entangled Photon Pair Source}
We use a type II colinear spontaneous-parametric-down-conversion (SPDC) process in a 1mm long periodically poled KTP crystal to create photon pairs entangled in their orbital-angular-momentum (OAM) degree of freedom. Due to the conservation of OAM in the SPDC process the created quantum state reads $c_0 \ket{0,0}+c_1 \ket{1,-1}+c_1 \ket{-1,1}+c_2 \ket{2,-2}+\cdots$, where the $|c_i|^2$ coefficients give the probability that certain OAM modes are created. Naturally it is more likely that an OAM mode $\ell=0$ is created than a $\ell=2$ mode, for example. But the distribution of the $c_i$ coefficients can be tuned to some extend using the pump beam and the collection beam waist~\cite{miatto2012bounds}. Here the pump beam waist is set to $35\mu m$ and the collection beam waist to $22\mu m$. This results in an unequal distribution of the $c_i$ coefficients. The ratio between the $c_0:c_1\approx 1.7:1$ and $c_1:c_2\approx 2:1$. We chose this unequal distribution because for an ideal equal distribution the absolut count rates in the $\ell=0$ and $\ell=\pm 1$ mode would be lower. And in our setup these are the only modes that are required to generate the three-dimensional GHZ state. 

The SPDC process is driven by a pulsed laser centered at $404\text{nm}$ with a pulse length of $\approx 140\text{fs}$, a repetition rate of 80~MHz and a maximum power of 700~mW. This results in approximately 13.000 photon pairs in the $\ket{0,0}$ and 4.100 photon pairs in the $\ket{\pm 1,\mp 1}$ mode per second and per nm spectral bandwidth.

\subsection{Temporal Distinguishability of Photon-Pairs Created in Separate non-linear Crystals}
The interference visibility of two photon pairs created in a SPDC process at a beam splitter critically depends on the temporal distinguishability of these photon pairs. In the analysis given here we closely follow~\cite{ou2007multi}. For simplicity we model our system as two non-linear crystals (ppKTP) phase-matched for Type-II down-conversion, where the two signal photons are overlapped at a beam splitter. Thus equation (8.18) from~\cite{ou2007multi} 
\begin{eqnarray*}
	P_4(\Delta T)=C \int d\omega_1 d\omega '_1 d\omega_2 d\omega '_2\\|\phi(\omega_1,\omega_2)\phi(\omega '_1,\omega '_2)-\phi(\omega '_1,\omega_2)\phi(\omega_1,\omega '_2) e^{i(\omega '_1-\omega_1)\Delta T}|^2,
\end{eqnarray*}
describing the probability of detecting a four-photon ``click'' event applies, where $C$ is a normalization constant, $\Delta T$ the time difference between the photons arriving at the beam splitter and $\phi(\omega_i,\omega_j)$ describing the joint-spectral-amplitude (JSA) of the SPDC process. A necessary and sufficient criterion to obtain the maximum visibility of one at $\Delta T\rightarrow 0$ is that the JSA is symmetric under exchange of the arguments and also factorisable $\phi(\omega_i,\omega_j)=\phi(\omega_j,\omega_i)=\psi(\omega_i)\psi(\omega_j)$. The symmetry of the JSA is dependent on the group velocity of the signal and idler photons. Since we are using a Type-II SPDC source, these group velocities are different. Nevertheless, it turns out that in our case the factorize-ability sets significantly more stringent constraints on the spectral filtering. Therefore, the difference in the group velocities for horizontally and vertically polarized photons is neglected. To gain further physical insight in the factorize-ability condition on the JSA function, we use a simplified model $\phi(\omega_i,\omega_j)=\alpha(\omega_i, \sigma_f)\alpha(\omega_j,\sigma_f)\beta(\omega_i,\omega_j,\Delta k)$, with $\alpha(\omega_i, \sigma_f)=\text{Exp}(-(\omega_i/ \sigma_f)^2)$ describing a Gaussian spectral filter and $\beta(\omega_i,\omega_j,\Delta k)=\gamma(\omega_p, \sigma_f)\text{sinc}(\Delta k(\omega_i,\omega_j) L/2)$ with $\Delta k(\omega_i,\omega_j)$ characterizing the phase-matching condition, $L$ is the crystal length and $\gamma(\omega_p, \sigma_p)=\text{Exp}(-(\omega_p/ \sigma_p)^2)$ modeling the spectral distribution of the pump beam. For a femto-second pulsed laser, 150fs in our case, the pump spectral width is Fourier limited to approximately $\sigma_p\approx 2\text{nm}$ at 404nm center wavelength. The phase-matching condition for a type II collinear SPDC process can be approximated to first order with $\Delta k(\omega_i, \omega_j)\approx (v_p^{-1}-v_i^{-1})\omega_i+(v_p^{-1}-v_s^{-1})\omega_s$ with $v_i=v_s$ denoting the group-velocity of the respective wavelength and polarization (which is neglected here). For simplicity we approximate the sinc-function with a Gaussian-function of approximately the same full width at half maximum value. This results in a new JSA given by $\phi'(\omega_i,\omega_j)=\alpha(\omega_i, \sigma_f)\alpha(\omega_j,\sigma_f)\beta'(\omega_i,\omega_j,\Delta k)$, with $\beta'(\omega_i,\omega_j,\Delta k)=\gamma(\omega_p, \sigma_f)\text{Exp}(-(\omega_i+\omega_j)^2/(\sigma_{GVM}^2)$, with $\sigma_{GVM}=2/(\sqrt{5}L(v_p^{-1}-v_{i,j}^{-1}))$. The factorize-ability of the JSA is crucially dependent on $\sigma_{GVM}$, which denotes the width according to the group velocity mismatch and the length of the crystal. In order to obtain a factorisable JSA, the spectral filtering $\sigma_f$ needs to be smaller or at least equal to $\sigma_{GVM}$. The visibility is now defined as $V=\frac{P_4(\Delta T\rightarrow\infty)-P_4(\Delta T\rightarrow0)}{P_4(\Delta T\rightarrow\infty)}$. Calculating the visibility with the above mentioned approximations leads to 
\begin{equation*}
	V=\frac{\sigma_{GVM}\sqrt{2 \sigma_f^2+\sigma^2_{GVM}}}{\sigma_f^2+\sigma^2_{GVM}}.
\end{equation*}
This formula underpins the physical intuition that two pairs of photons created in different crystals are temporal indistinguishable if the possible timing resolution due to their spectral bandwidth is larger then the average timing jitter in their creation time. This average timing jitter is governed by the group velocity mismatch between pump and down-converted photons, as well as the length of the crystal.
For a 1mm long ppKTP crystal $\sigma_{GVM}\approx 559 \text{GHz}$ which corresponds to $\approx 1.2\text{nm}$ at a center-wavelength of $808\text{nm}$.
Thus we chose our filters to be at the same value of $\sigma_f=1.2\text{nm}$ and expect therefore a visibility of approximately 86\%.
\subsection{Detailed Calculation of $\rho_p$}
\begin{equation}
	\rho_{\text{p}}=p \ket{\text{GHZ}}\bra{\text{GHZ}}+\frac{(1-p)}{27} \mathds{1}_{27}-c\rho_{\text{dep.}},
\end{equation}
with $\ket{\text{GHZ}}$ denoting the ideal GHZ state $\ket{\text{GHZ}}=\alpha\ket{000}+\beta\ket{111}+\gamma\ket{222}$, $\mathds{1}_{27}$ representing a 27-dimensional identity matrix simulating white-noise $p$ and $\rho_{\text{dep.}}$ denoting the dephasing term simulating the average coherence $c$.

\subsection{Estimation of the generalized Mermin operator $\mathcal{O}$}
\begin{equation}
	\avg{\mathcal{O}}_{\rho_{\text{p}}}=\frac{9 c p \left(\alpha  \beta ^*+\gamma  \alpha ^*+\beta  \gamma ^*\right)}{\left| \alpha \right| ^2+\left| \beta \right| ^2+\left| \gamma \right| ^2}=6.26\pm0.25
\end{equation}
Estimated values for calcuting the generalized Mermin operator from experimentally measured density matrix.

\begin{table}[ht]
\centering
\caption{Estimated values for calculating the generalized Mermin operator from experimentally measured density matrix.}
\label{tab:table-1}
\begin{tabular}{|c|c|c|c|c|}
\hline
White-Noise {[}p{]} & avg. coherence & $\alpha$ & $\beta$ & $\gamma$ \\ \hline
0.878               & 0.817          & 0.685    & 0.588   & 0.491    \\ \hline
\end{tabular}
\end{table}

\subsection{Higher-Order Effects}
In our experimental setup there are two sources of higher-order contributions. First, due to the natural conservation of OAM in the SPDC process the overall state produced by our SPDC source reads $c_0 \ket{0,0}+c_1 \ket{1,-1}+c_1 \ket{-1,1}+c_2 \ket{2,-2}+\cdots$. The higher OAM values $\ell=2,3,4,...$ are unwanted terms in our setup. Fortunately, all these terms don't contribute to the final state. This stems from the fact that we only take events into account where all four detectors simultaneously detect a photon and detector A can only detects either $\ell=0$ or $\ell=-1$ photons. Thus we only need to check the probability amplitudes that can lead to these OAM quanta in detector A. Only photons from crystal 1 in path A undergo a reflection followed by a SPP which adds an OAM quanta of $+2$. All the other paths (B,C) before the multi-port do not change their absolute OAM quanta. Thus the probability amplitude $\ket{2,-2}_{AB}$ from crystal 1 is transformed to $\ket{0,-2}_{AB}$ and could be detected in detector A. This probability amplitude is not in the subspace for photon A, which is given by $\{-1,2,3\}$. Therefore it will not contribute to the final state. 

The second source of higher-order effects are double pair emissions of one crystal. If only one of the two non-linear crystals emits two pairs of photons, no simultaneous detection in all four detectors can occur. Thus we can focus on the case where one non-linear crystal emits one pair and the other emits two pairs. Let us look at the worst case scenario, which is that both crystals emit the $\ket{0,0}$ mode (which is the most likely one). The observed photon pairs per second are proportional to the repetition rate R of the laser, the overall detection efficiency $\eta$ squared and the mean photon number per pulse $\mu$. We estimate the overall detection efficiency (detector $\approx 65\%$, SLM $\approx 85\%$, SMF $\approx 80\%$) to be $44\%$. Thus for $13.000$ photon pairs per second we calculate a mean photon number per pulse of $\mu=8.4\times 10^{-4}$. There are three different possibilities in our setup that allow simultaneous four-photon detection events that could occur from six photon emission. The first two are that one crystal emits two pairs and the other crystal one pair, while the second possibility is that the second crystal emits 3 pairs simultaneously. Thus the ratio between a simultaneous six photon detection event and a four photon detection event is $\frac{3\times\mu^3\eta^6}{\mu^2\eta^4}=3\times\mu\eta^2=4.8\times 10^{-4}$. This is approximately 3 orders of magnitude lower than our expected four photon detection events. Therefore we can safely neglect higher order contributions from double pair emission.

\subsection{Prediction and Accidental-Count Subtraction of Simultaneous Four-Photon Detection Events}

A simultaneous four-photon detection event occurs purely statistically. The probability of two independent events to occur simultaneously is given by the product of the probabilities. Whenever crystal one and two simultaneously emit a pair of photons they can either arrive at the detector pair (AB)\&(CD), (AC)\&(BD) or (AD)\&(BC). The probability to observe such a four-photon event $p_\text{4-ph}$ is then given by the sum of the products of the probabilities $p_\text{i,j}$ that a photon pair is detected:
\begin{equation}\label{eq:theoretically-expected-4-folds-due-to-2-folds}
 	p_\text{4-ph}=p_\text{AB}\times p_\text{CD}+p_\text{AC}\times p_\text{BD}+p_\text{AD}\times p_\text{BC}.
 \end{equation} 
 \begin{figure}[htbp]
 	\includegraphics[width=0.45\textwidth]{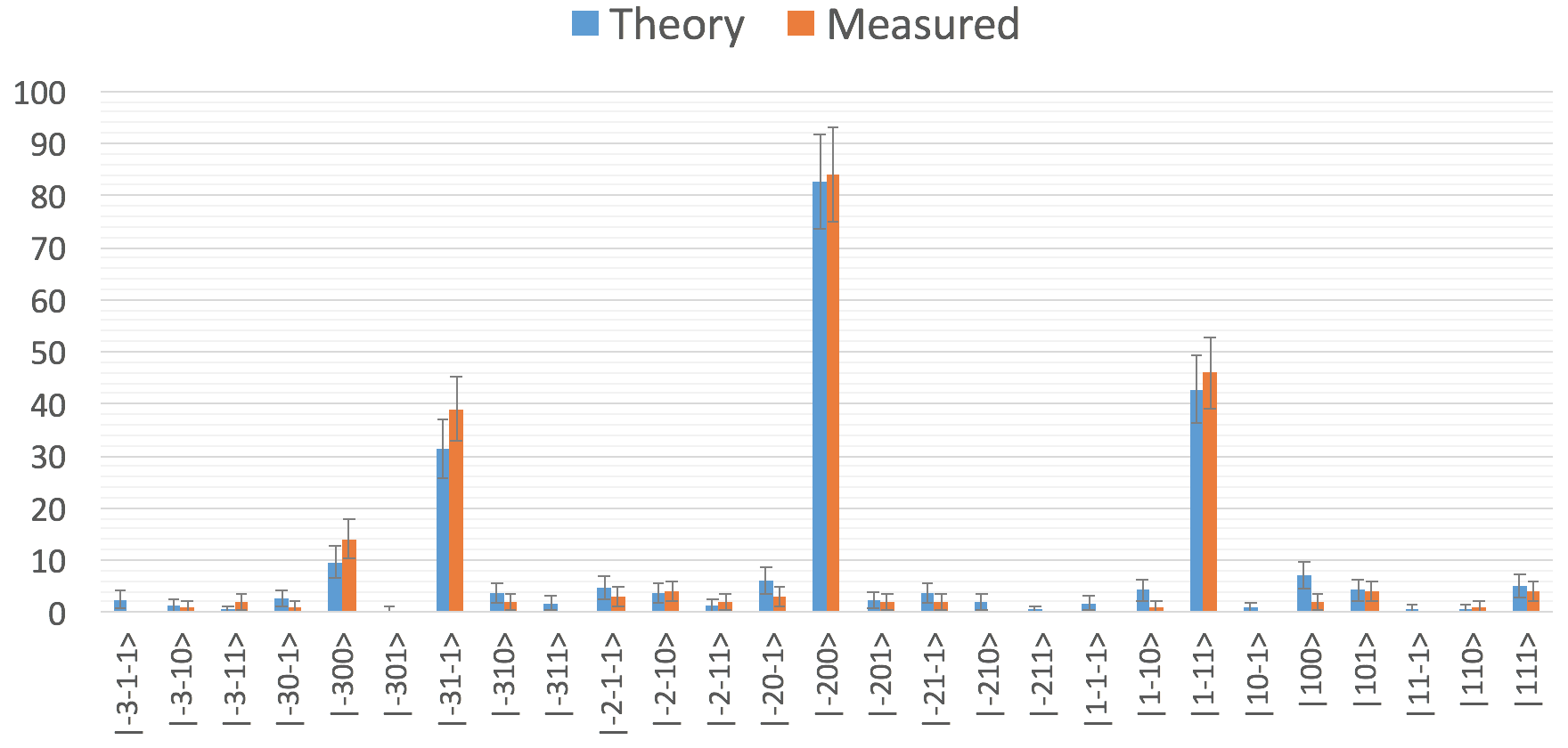}
 	\caption{\textbf{Comparison of expected and measured four-photon events.} Here we compare the theoretically expected four-photon events due to measured two-photon events according to equation~\ref{eq:theoretically-expected-4-folds-due-to-2-folds} with the actually measured four-photon events. Here all observed two-photon and four-photon events are taken into account. Error bars are given due to Poissonian counting statistics.}
 	\label{fig:exp_FF}
 \end{figure}
Now by measuring the actual photon pairs detected between the different detectors, we can predict how many simultaneous four-photon events we expect. A comparison between theory and our final measurements is shown in Fig.~\ref{fig:exp_FF} and it shows a very good overlap between theory and experiment.

Here we would like to discuss the issue of accidental four-photon detection and their origin in detail. As we have seen, the overall probability to observe a four-photon event is dependent on all three possible combinations of two photon events. Now the problem can be illustrated with the following example. We expect two photon events between detector C and D only in the $\ket{0,0}_\text{CD}$ state. Thus probing detectors C and D in the state $\ket{1,0}_\text{CD}$ should yield zero two photon events. But experimentally we still find a small probability for such events. This is because there is another expected two-photon event between detector B and C for the $\ket{-3,1}_\text{BC}$ state. This means there are independent single photon events between detector C and D in the state $\ket{1,0}_\text{CD}$. The probability for such an event to occur can be calculated by
\begin{equation}\label{eq:accidental-counts-equation-2f}
	\text{acc}_{i,j}=\frac{S_i\times S_j}{\tau_\text{int}^2\times \text{rep. Rate}}
\end{equation}
with $\tau_\text{int}$ denoting the integration time and rep. Rate is the repetition rate of the pulsed laser (in our case 80MHz). 
\begin{figure}[htbp]
	\includegraphics[width=0.45\textwidth]{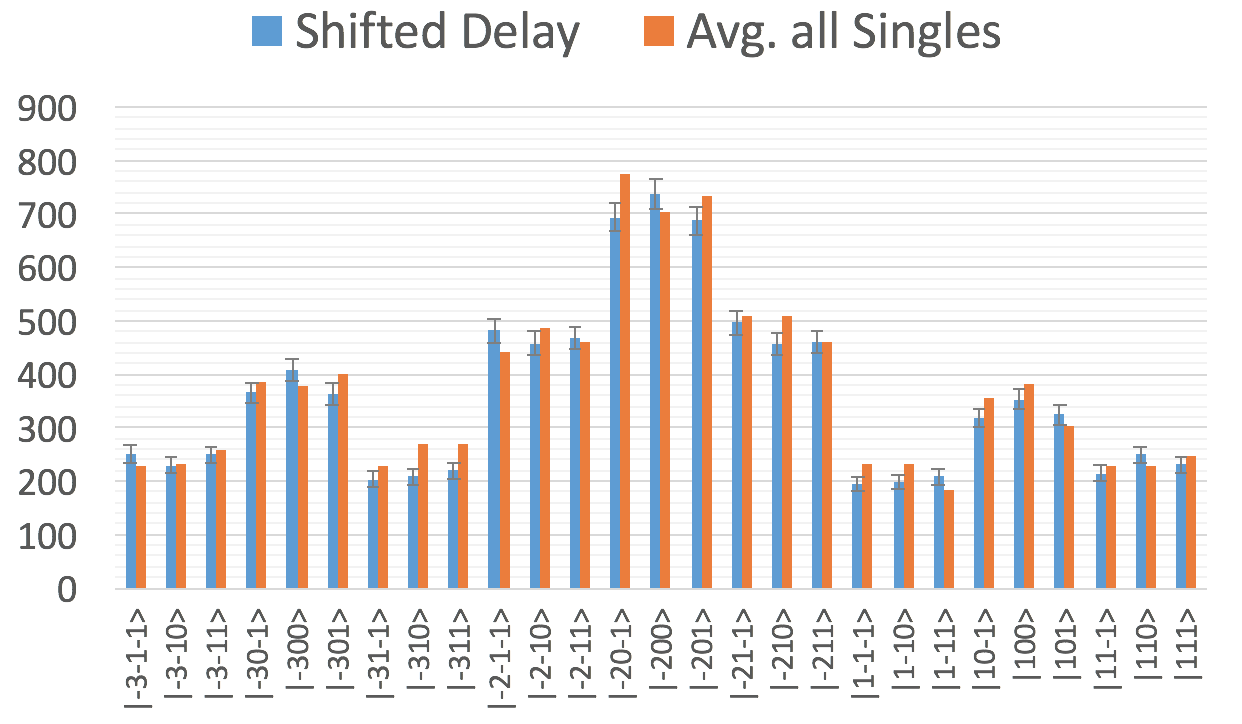}
	\caption{\textbf{Comparison of the actually measured accidentals and expected accidentals calculated from single photon events.} By shifting two detectors relative by one pulse (12.5ns) to each other we can actually measure the detected accidental counts that stem from single-photon events only. We compare this in this graph with the estimation from equation~\ref{eq:accidental-counts-equation-2f} and use the average of all single photon counts detected in the complete experiment. Error bars are due to Poissonian counting statistics.  }
	\label{fig:2F_shift}
\end{figure}
The validity of this approach can be checked experimentally. In order to do so we shift two detectors exactly by one pulse length (12.5 ns) with respect to each other. Thus two-photon events can only occur statistically any more. Fig.~\ref{fig:2F_shift} shows a comparison between the measurements where the two detectors are shifted relative to each other and the estimated accidental counts. The very good overlap between the measured and expected accidentals shown in Fig.~\ref{fig:2F_shift} validates our approach. These accidental two-photon events now contribute to the overall four-photon events in the following way:
\begin{eqnarray}\label{eq:accidental-counts-equation-4f}
	\text{acc}_\text{4-ph}&=\text{acc}_{AB}\times \text{CC}_{CD}+\text{acc}_{CD}\times \text{CC}_{AB}\\\nonumber
	&+\text{acc}_{AC}\times \text{CC}_{BD}+\text{acc}_{BD}\times \text{CC}_{AC}\\\nonumber
	&+\text{acc}_{AD}\times \text{CC}_{BC}+\text{acc}_{BC}\times \text{CC}_{AD},
\end{eqnarray}
with $\text{CC}_{i,j}$ denoting the two-photon events between detectors $i$ and $j$. In Fig.~\ref{fig:sub_FF} we show the accidental four-photon events that we subtract from the actually observed ones. The resulting density matrix where the accidental four-photon events have been subtracted is then used to calculate the entanglement witness and to estimate the expectation value for the Mermin operator.
\begin{figure}[htbp]
	\includegraphics[width=0.45\textwidth]{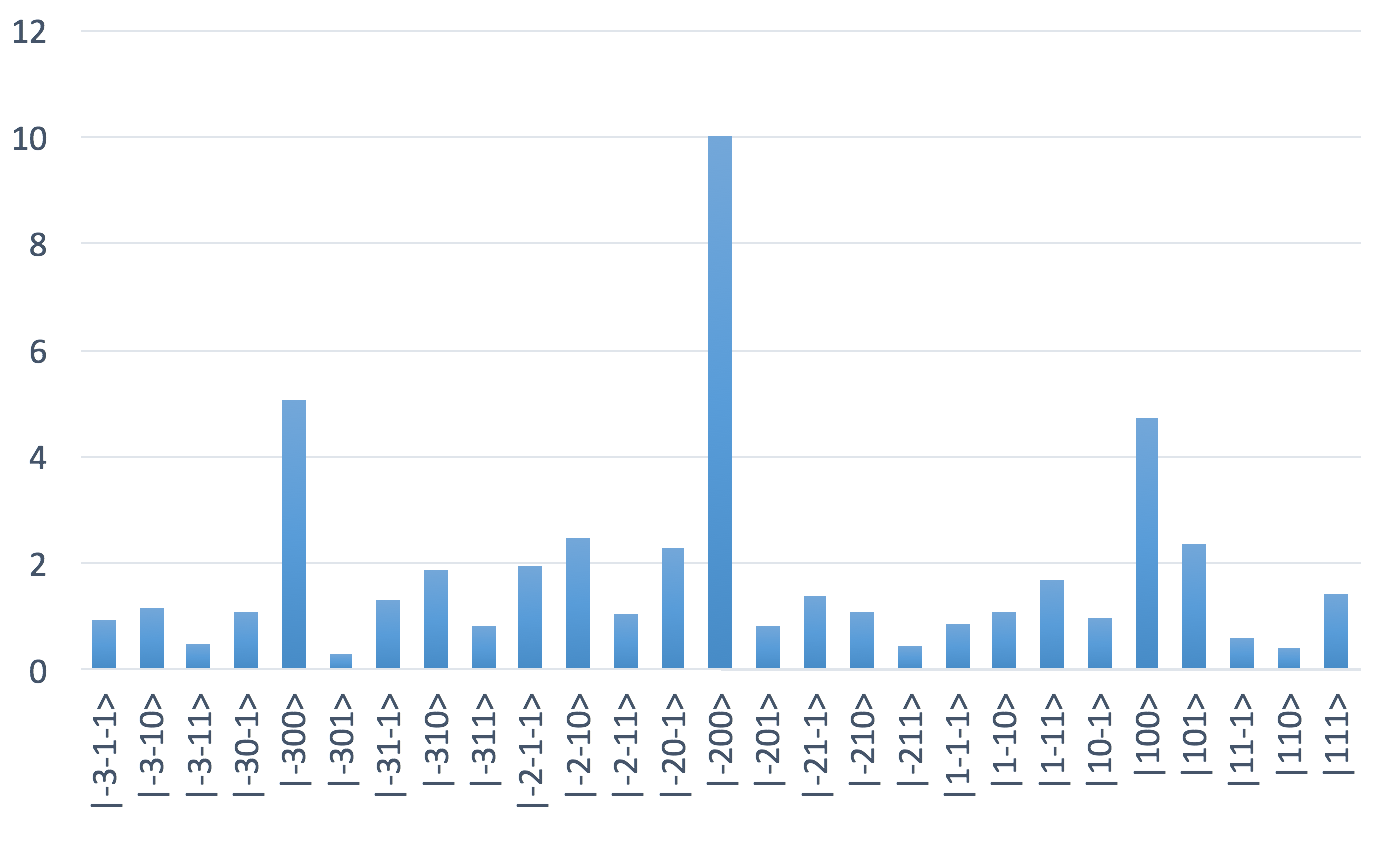}
	\caption{\textbf{Calculated accidental four-photon events.} The calculated accidental four-photon events from equation~\ref{eq:accidental-counts-equation-4f} are shown for the diagonal elements. It is clearly visible, that whenever the $\ket{0,0}$ mode from crystal two is involed, the accidental four-photon rate is significantly higher than the others. }
	\label{fig:sub_FF}
\end{figure}

\subsection{Detailed Experimental Setup}

The experimental setup depicted below includes additional mirrors that are not shown in Fig.~1 for simplicity. With these elements included, our state is unitarily modified to the following form:
\begin{equation}
	\ket{\psi}=\frac{1}{\sqrt{3}}\left(\ket{-2,0,0}+\ket{-3,1,-1}+\ket{1,-1,1}\right)
\end{equation}
\begin{figure*}[ht]
	\centering
	\includegraphics[width=0.95\textwidth]{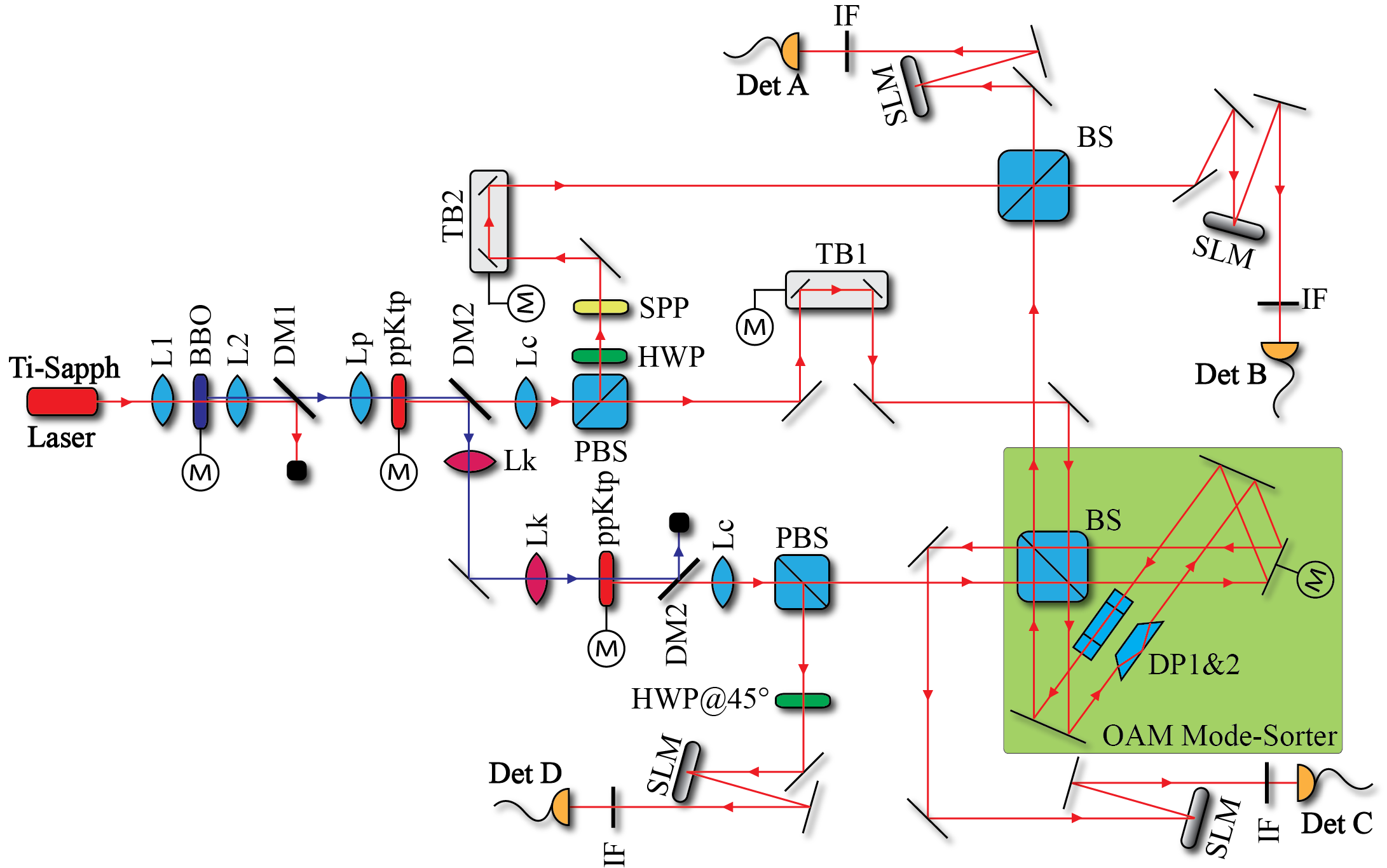}
	\caption{\textbf{Detailed Experimental Setup.} A Ti:Sapphire pulsed laser source with pulse duration of 140fs and repetition rate of 80 MHz centered at 808nm is focused by lens L1 into a $\beta$-Barium Borate crystal (BBO) to produce pump pulses at 404nm via the process of second harmonic generation. In order to avoid damage, the BBO crystal is moving periodically with a motor (M). Lens L2 re-collimates the 404nm pump pulses, which are separated from the 808nm laser by a dichroic mirror (DM1). Two OAM-entangled photon pairs are produced via type-II spontaneous parametric down-conversion in two periodically poled Potassium Titanyl Phosphate (ppKTP) crystals (NL1 and NL2) of dimensions $1\times 2\times 1$mm. The full-width half-maximum beam waist of the 404nm pump laser at the crystals is 35$\mu$m with an average power of 700mW. To prevent gray-tracking and crystal damage, the ppKTP crystals are moved periodically up and down in steps of 200$\mu$m via motors (M). A second dichroic mirror (DM2) separates the entangled photon pairs from the 404nm pump laser beam. Lenses Lc are used to collimate the down-converted 808nm photon pairs. A $4f$ imaging system Lk is used to perfectly image the 404nm pump beam from crystal NL1 to crystal NL2 to ensure that the pump beam mode is exactly the same at both crystals. The two lenses Lk are chosen such that they compensate for the Kerr-lensing effect in the first non-linear crystal. Polarizing beam splitter (PBS) are used to deterministically separate all four photons. A moving trombone system of mirrors (TB1) is used to ensure that photons from both crystals arrive at the OAM mode sorter at the same time.	A second moving trombone system of mirrors (TB2) is used to ensure that photons from the first crystal arrive at the beam-splitter (BS) at the same time. Thus it is guaranteed that all photons emitted from the first NLC and the second NLC can interfere coherently with each other. The OAM beam splitter is implemented in a double Sagnac interferometer configuration that allows for stable operation over several days. Note that this is different from the physical implementation shown in Fig.~\ref{fig:exp-scheme} (for simplicity), but has exactly the same physical outcome. A piezo controlled mirror (M) is scanned every hour to optimize the alignment of the interferometer and ensure that it is sorting OAM modes correctly. Two half-wave plates (HWPs) at 45$^\circ$ rotate the photon polarization from vertical to horizontal in paths A and D. Projective measurements are performed  by four spatial-light-modulators (SLMs) in combination with aspheric lenses and single mode fibers, which guide the photons to single photon avalanche diodes (Det A, B, C, and D). Additionally, narrow-band filters (IF) with 1.2nm FWHM spectral width at 808nm are used before all four detectors to ensure spectral indistinguishability between the interfering photons.}
	\label{fig:detailed-exp-setup}
\end{figure*}

\subsection{$\mathbf{(333)}$-Entanglement witness} 
In order to prove that the state is indeed a $(333)$-type entangled state we have to prove that it cannot be decomposed into states of a smaller dimensionality structure. We thus have to show that it lies outside the $(332)$-set of states, that is the convex hull of all states that can be decomposed into $322,323$ and $233$ states. This witness was developed in~\cite{malik2016multi} and we follow this approach here. From the measured data we can extract the fidelity to the ideal state
\be
	\ket{\Psi}=\frac{1}{\sqrt{3}}\left( \ket{2,0,0}+\ket{-1,-1,-1}+\ket{3,1,1} \right)\,,
	\label{ideal}
\ee
which we will denote as $F_\text{exp}:=\text{Tr}(\rho_\text{exp}|\Psi\rangle\langle\Psi|)$. We thus need to compare the experimental fidelity with the best achievable fidelity of a $(332)$-state, i.e.
\begin{align}
F_\text{max}:=\max_{\sigma\in (332)}\text{Tr}(\sigma |\Psi\rangle\langle\Psi|)\,.
\end{align}
If $F_\text{exp}>F_\text{max}$, we can conclude that the experimentally certified fidelity cannot be explained by any state in $(332)$ and thus the underlying state is certified to have an entangled dimensionality structure of $(333)$. In~\cite{malik2016multi} it is shown how to calculate $F_\text{max}$. According to Eq.~(11, 12 and 13) in~\cite{malik2016multi} we only need to calculate the Schmidt coefficients $\lambda_i$ of our target state $\ket{\Psi}$ for all three possible bi-partitions A|BC, B|AC and C|AB:
\begin{align}
	\text{they are}~~\{1/\sqrt{3},~1/\sqrt{3},~1/\sqrt{3}\}.
\end{align}
Now due to equation~(11 and 12) in~\cite{malik2016multi} $F_\text{max}$ is given by the sum of all but the smallest Schmidt coefficients squared, which is in our case
\begin{align}
F_\text{max}=\frac{1}{3}+\frac{1}{3}=\frac{2}{3}\,.
\end{align}

\subsection{Witness Measurements} 
The experimental fidelity $F_{\text{exp}}:=\text{Tr}(\rho_{\text{exp}}|\Psi\rangle\langle\Psi|)$ determines which measurements are required. The projector $|\Psi\rangle\langle\Psi|$ projects only onto the non-zero diagonal and off-diagonal elements contained in the density matrix $\rho_{\text{exp}}$. Additionally, for the purpose of normalization, it is necessary to measure all other diagonal elements in $\rho_{\text{exp}}$. This results in 27 diagonal and 3 unique off-diagonal elements that need to be measured in order to calculate $F_{\text{exp}}$. In our experiment, we can only perform projective measurements with SLMs. A diagonal element is given by one single projection $\langle ijk|\rho|ijk\rangle=\frac{C(ijk)}{C_T}$, with $C_T:=\sum_{i=-1,0,1}\sum_{j=-1,0,1}\sum_{k=0,1}C(ijk)$ containing all diagonal elements for normalization. Out of the six off-diagonal elements, only three are unique and need to be measured: $\langle 000|\rho|1\text{-}11\rangle$, $\langle 000|\rho|\text{-}111\rangle$ and $\langle \text{-}111|\rho|1\text{-}11\rangle$. Note that the last off-diagonal element is only in a two-particle superposition. Hence, it can be measured in the standard way that two-particle two-dimensional states are usually measured. In order to measure the other two off-diagonal elements with projective measurements, we decompose them into $\sigma_\text{x}$ and $\sigma_\text{y}$ measurements. The real and imaginary part of each element can be written as 
\begin{widetext}
\begin{eqnarray}\label{eq:real-imaginary-off-diagonals}
\Re\big[ \langle ijk|\rho|lmn\rangle \big]=\expval{\sig{x}{i}{l}\otimes\sig{x}{j}{m}\otimes\sig{x}{k}{n}}-\expval{\sig{y}{i}{l}\otimes\sig{y}{j}{m}\otimes\sig{x}{k}{n}}-\expval{\sig{y}{i}{l}\otimes\sig{x}{j}{m}\otimes\sig{y}{k}{n}}-\expval{\sig{x}{i}{l}\otimes\sig{y}{j}{m}\otimes\sig{y}{k}{n}}\\\nonumber
\Im\big[ \langle ijk|\rho|lmn\rangle \big]=\expval{\sig{y}{i}{l}\otimes\sig{y}{j}{m}\otimes\sig{y}{k}{n}}-\expval{\sig{x}{i}{l}\otimes\sig{x}{j}{m}\otimes\sig{y}{k}{n}}-\expval{\sig{x}{i}{l}\otimes\sig{y}{j}{m}\otimes\sig{x}{k}{n}}-\expval{\sig{y}{i}{l}\otimes\sig{x}{j}{m}\otimes\sig{x}{k}{n}},
\end{eqnarray}
\end{widetext}
where $ \sig{x}{a}{b}=\ket{a}\bra{b}+\ket{b}\bra{a} $ and $ \sig{y}{a}{b}=i \ket{a}\bra{b} - i \ket{b}\bra{a} $. The $\sigma_{x,y}$ operators are also not measurable directly with SLMs and are therefore rewritten using the following operators:
\begin{widetext}
\begin{eqnarray}
	\widehat{\mathcal{P}}_+(a,b)=\ket{+}\bra{+}_{(a,b)}=\ket{a}\bra{a}+\ket{b}\bra{b}+\ket{a}\bra{b}+\ket{b}\bra{a}\\\nonumber
	\widehat{\mathcal{P}}_-(a,b)=\ket{-}\bra{-}_{(a,b)}=\ket{a}\bra{a}+\ket{b}\bra{b}-\ket{a}\bra{b}-\ket{b}\bra{a}\\\nonumber
	\widehat{\mathcal{P}}_{+i}(a,b)=\ket{+i}\bra{+i}_{(a,b)}=\ket{a}\bra{a}+\ket{b}\bra{b}-i\ket{a}\bra{b}+i\ket{b}\bra{a}\\\nonumber
	\widehat{\mathcal{P}}_{-i}(a,b)=\ket{-i}\bra{-i}_{(a,b)}=\ket{a}\bra{a}+\ket{b}\bra{b}+i\ket{a}\bra{b}-i\ket{b}\bra{a},
\end{eqnarray}
\end{widetext}
where $\ket{+}_{(a,b)}=\ket{a}+\ket{b}$, $\ket{-}_{(a,b)}=\ket{a}-\ket{b}$, $\ket{+i}_{(a,b)}=\ket{a}+i\ket{b}$ and $\ket{-i}_{(a,b)}=\ket{a}-i\ket{b}$. These superposition states can be measured with SLMs in our experiment. Thus the $\sigma$ operators from Eq.~\ref{eq:real-imaginary-off-diagonals} can be written in the following manner:
\begin{eqnarray}
	\sig{x}{a}{b}=\frac{1}{2}\left( \widehat{\mathcal{P}}_+(a,b) -\widehat{\mathcal{P}}_-(a,b) \right)\nonumber\\
	\sig{y}{a}{b}=\frac{1}{2}\left( \widehat{\mathcal{P}}_{-i}(a,b) -\widehat{\mathcal{P}}_{+i}(a,b) \right).
\end{eqnarray}
This leads to 64 projection measurements required for measuring each of the three off-diagonal elements. Summing up these measurements as well as the 27 diagonal elements leads to 219 total measurements. From these measurements, the overlap between the generated state $\rho_\text{exp}$ and the ideal state $(332)$-state $\ket{\Psi}$ is calculated to be $F_\text{exp}=75.2\%\pm2.88\%$. The error in the overlap is calculated by propagating the Poissonian error in the photon-counting rates by performing a Monte Carlo simulation of the experiment.

\subsection{Details of the Three-Dimensional Multi-Setting GHZ Argument}
The three-dimensional cyclic-ladder operator X is defined as $X\ket{\ell}=\ket{\ell\oplus 1}$. The other two local observables are rotated cyclic operators $Y=Z^{1/3}XZ^{-1/3}$ and $W=Z^{2/3}XZ^{-2/3}$, with $Z\ket{\ell}=\omega^\ell\ket{\ell}$ and $\omega=e^{2\pi i/3}$. The possible outcomes of these local observables are given by their set of eigenvalues $\{1,\omega, \omega^2\}$.

Next, we choose a set of three-body observables with the GHZ state as eigenstate. The following concurrent set leads to a genuine three-dimensional GHZ contraction:
\begin{align*}
\Big\{XXX~(1), ~YYY~(\omega),~WWW~(\omega^2),\\~XYW~(\omega),~XWY~(\omega),~YXW~(\omega),\\~YWX~(\omega),~WXY~(\omega),~WYX~(\omega)\Big\},
\end{align*}
where we denote the quantum mechanical expectation value for the GHZ state in brackets $(\cdot)$. Next, we define a generalized Mermin operator such that it maximizes the quantum mechanical expectation value:
\begin{align*}
	\mathcal{O}=XXX+\omega^{-1} YYY+\omega^{-2} WWW\\+\omega^{-1}\Big(XYW+XWY+YXW\\+YWX+WXY+WYX\Big).
\end{align*}
 Since every single term in this sum has an expectation value equal to one, the expectation value is $\langle \mathcal{O}\rangle_{\text{GHZ}}=9$. If we now assign values according to a local-realistic theory, each single particle observable $A_k$ takes a definite value $v(A_k)\in \{ 1,\omega,\omega^2\}$ in 
 \begin{align*}
 	X_1X_2X_3+\omega^{-1} Y_1Y_2Y_3+\omega^{-2} W_1W_2W_3\\+\omega^{-1}\Big(X_1Y_2W_3+X_1W_2Y_3+Y_1X_2W_3\\+Y_1W_2X_3+W_1X_2Y_3+W_1Y_2X_3\Big).
 \end{align*}
The maximum value out of all $3^9=19683$ possibilities under these local-realistic constraints is given by $6$.

\end{document}